\documentstyle[aps,epsf]{revtex} % PH. REV. FINAL FORMAT STYLE
\begin{document}

\twocolumn[\hsize\textwidth\columnwidth\hsize\csname@twocolumnfalse\endcsname

\title{Dynamics of collapsing and exploding Bose-Einstein condensate}

\author{Sadhan K. Adhikari}
\address{Instituto de F\'{\i}sica Te\'orica, Universidade Estadual
Paulista, 01.405-900 S\~ao Paulo, S\~ao Paulo, Brazil\\}

\date{\today}
\maketitle
\begin{abstract}

Recently, Donley et al. performed an experiment on the
dynamics of collapsing and exploding Bose-Einstein condensates by suddenly
changing the scattering length of atomic interaction to a large
negative value  on a preformed repulsive
condensate of  $^{85}$Rb atoms in an axially symmetric 
trap.  Consequently, the condensate collapses and ejects atoms via  
explosions.
We show that the accurate numerical solution of the time-dependent
Gross-Pitaevskii equation with axial symmetry can explain some aspects
of the dynamics of the collapsing condensate.

%{\bf PACS Number(s):  03.75.Fi}

\end{abstract}

\vskip1.5pc]
 \newpage

Since the successful detection \cite{1a,1b,1c,ex2a,ex2b} of Bose-Einstein
condensates
(BEC) in dilute bosonic atoms employing magnetic trap at ultra-low
temperature, one problem of extreme interest is the dynamical
study of the formation and decay of BEC for attractive atomic interaction
\cite{ex2a,ex2b}.

For attractive  interaction  the condensate is
stable for a maximum critical number $N_{\mbox{cr}}$ of atoms
\cite{ex2a}.  When the
number of atoms
increases beyond this critical number, due to interatomic attraction the
radius of BEC tends to zero and the central density tends to infinity.
Consequently, the condensate collapses emitting atoms until the number of
atoms is reduced below $N_{\mbox{cr}}$  and a stable configuration is
reached. With a supply of atoms from an external source the condensate can
grow again and thus a series of collapses can take place, which was
observed experimentally in the BEC of $^7$Li with attractive interaction
\cite{ex2a}. Theoretical analyses based on the 
time-dependent mean-field Gross-Pitaevskii (GP) 
equation also confirm
the collapse \cite{th1a,th1b,th2a,th2b,th2c,11}.

It is possible to manipulate the interatomic interaction
by an external magnetic field
via a Feshbach resonance \cite{fs}. Roberts et al. \cite{ex3a}
and Cornish et
al. \cite{ex3b} (at JILA)
 exploited this idea
to
suddenly change the atomic scattering length by a large amount in
experiment. They have been able to even change the sign
of the
scattering length, thus changing a repulsive condensate in to an
attractive one and {\it vice versa}. Consequently, a stable preformed
repulsive condensate is
suddenly turned to a highly explosive and collapsing attractive
condensate.
In a classic experiment performed at JILA
Donley et al. \cite{ex4}  studied the dynamics of
collapsing and
exploding  condensates formed of $^{85}$Rb atoms. The natural
scattering
length of  $^{85}$Rb atoms is negative (attractive). By exploiting the
Feshbach
resonance they made it positive (repulsive) in the initial state.
They have provided a
quantitative measurement of this explosion by counting the number of
emitted and remaining 
atoms
in the condensate as a function of time until an equilibrium is
reached.  They also measured the energy distribution of the emitted atoms.
They claim that their experiment reveal many interesting  
phenomena that challenge theoretical models.

We demonstrate that some aspects  of the
above collapse and explosion of the attractive condensate of $^{85}$Rb
atoms can be explained from an accurate numerical solution of the
GP  equation  in  an
axially symmetric trap, where we include a quintic three-body
nonlinear recombination term which accounts for the decay of the strongly
attractive condensate. 
The  numerical method, we use,  for the solution of the
time-dependent GP equation with an axially symmetric trap has appeared
elsewhere \cite{sk1,sk2a,sk2b,sk2c1,sk2c,sk2d,sk2e}.

There have been other theoretical studies \cite{th1a,th1b,th2a,th2b,th2c}
to deal
with
dynamical
collapse including an absorptive term to account for the loss of
particles. Instead of attempting a full numerical solution of the GP
equation with axial symmetry, these investigations used various
approximations to study the time evolution of the condensate or employed
a spherically symmetric trap.

The time-dependent Bose-Einstein condensate wave
function $\Psi({\bf r};\tau)$ at position ${\bf r}$ and time $\tau $ may
be described by the following  mean-field nonlinear GP equation
\cite{11}
\begin{eqnarray}\label{a} \biggr[ -\frac{\hbar^2\nabla^2   }{2m}
+ V({\bf r})  
+ gN|\Psi({\bf
r};\tau)|^2
-i\hbar\frac{\partial
}{\partial \tau} \biggr]\Psi({\bf r};\tau)=0.  \nonumber \\ \end{eqnarray}
Here $m$
is
the mass and  $N$ the number of atoms in the
condensate, 
 $g=4\pi \hbar^2 a/m $ the strength of interatomic interaction, with
$a$ the atomic scattering length. 
The trap potential with cylindrical symmetry may be written as  $  V({\bf
r}) =\frac{1}{2}m \omega ^2(r^2+\lambda^2 z^2)$ where 
 $\omega$ is the angular frequency
in the radial direction $r$ and 
$\lambda \omega$ that in  the
axial direction $z$. We are using the cylindrical
coordinate system ${\bf r}\equiv (r,\theta,z)$ with $\theta$ the azimuthal 
angle.
The normalization condition of the wave
function is
$ \int d{\bf r} |\Psi({\bf r};\tau)|^2 = 1. $

In the absence of angular
momentum the wave function has the form $\Psi({\bf
r};\tau)=\psi(r,z;\tau).$
Now  transforming to 
dimensionless variables
defined by $x =\sqrt 2 r/l$,  $y=\sqrt 2 z/l$,   $t=\tau \omega, $ $l\equiv \sqrt {\hbar/(m\omega)}$, 
and  
\begin{equation}\label{wf}
\phi(x,y;t)\equiv 
\frac{ \varphi(x,y;t)}{x} =  \sqrt{\frac{l^3}{\sqrt 8}}\psi(r,z;\tau),
\end{equation} 
we get 
\begin{eqnarray}\label{d}
\biggr[-i\frac{\partial
}{\partial t}& -&\frac{\partial^2}{\partial
x^2}+\frac{1}{x}\frac{\partial}{\partial x} -\frac{\partial^2}{\partial
y^2}
+\frac{1}{4}\left(x^2+\lambda^2 y^2-\frac{4}{x^2}\right) \nonumber \\
&+& 8 \sqrt 2 \pi   n\left|\frac {\varphi({x,y};t)}{x}\right|^2 
 \biggr]\varphi({ x,y};t)=0, 
\end{eqnarray}
where
$ n =   N a /l.$ 
The normalization condition  of the wave
function becomes 
\begin{equation}\label{5} {\cal N}_{\mbox{norm}}\equiv {2\pi} \int_0
^\infty
dx \int _{-\infty}^\infty dy|\varphi(x,y;t)|
^2 x^{-1}=1.  \end{equation}
The root mean square (rms) sizes  $x_{\mbox{rms}}$ and  $y_{\mbox{rms}}$
are
defined by
\begin{eqnarray}
x^2_{\mbox{rms}}= {\cal N}_{\mbox{norm}}^{-1} {2\pi} \int_0
^\infty
dx \int _{-\infty}^\infty dy|\varphi(x,y;t)|
^2 x,   \\
y^2_{\mbox{rms}}= {\cal N}_{\mbox{norm}}^{-1} {2\pi} \int_0
^\infty
dx \int _{-\infty}^\infty dy|\varphi(x,y;t)|
^2 y^2x^{-1}.  
\end{eqnarray}

It is now appropriate to calculate the parameters of the present
dimensionless GP equation (\ref{d}) corresponding to the experiment
 at JILA \cite{ex4}. We follow the notation and nomenclature of
Ref. \cite{ex4}.
Their radial and axial trap frequencies are $\nu_{\mbox{radial}}=17.5$ Hz
and  $ \nu_{\mbox{axial}}=6.8$ Hz,
respectively, leading to $\lambda = 0.389 $. The harmonic oscillator
length $l$ for $^{85}$Rb atoms for $\omega =2\pi\times 17.5$ Hz is
$l=\sqrt{\hbar/(m\omega)}=25905$ \AA. One unit of time $t$ of
Eq. (\ref{d}) is $1/\omega$ or 0.009095 s. 
They prepared a stable $^{85}$Rb 
condensate of $N_0= 16000$ atoms with scattering  length
$a=a_{\mbox{initial}}=7a_0$,
$a_0=0.5292$ \AA, such that the initial $n=2.288$. Then during an
interval of time 0.1 ms the scattering length was ramped to  $a=
a_{\mbox{collapse}}=-30a_0$
such that final $n=-9.805$. The final condensate is strongly attractive
and unstable and undergoes a sequence of collapse and explosion.

The sequence of collapse in many theoretical studies has been explained by
introducing an absorptive imaginary  three-body quintic  interaction term
of strength $\xi$
responsible
for recombination loss.  Consequently Eq. (\ref{d}) becomes
\cite{th1a,th1b,th2a}
\begin{eqnarray}\label{d1}
\biggr[-i\frac{\partial
}{\partial t} -\frac{\partial^2}{\partial
x^2}+\frac{1}{x}\frac{\partial}{\partial x} -\frac{\partial^2}{\partial
y^2}
+\frac{1}{4}\left(x^2+\lambda^2 y^2-\frac{4}{x^2}\right) + \nonumber \\
 8 \sqrt 2 \pi   n\left|\frac {\varphi({x,y};t)}{x}\right|^2
- i\xi n^2\left|\frac {\varphi({x,y};t)}{x}\right|^4 
 \biggr]\varphi({ x,y};t)=0. 
\end{eqnarray}
There are many ways to account for the loss mechanism \cite{th1a,th1b}. It
is
quite impossible to include them all in a self consistent fashion. Here we 
simulate the loss via 
the most important quintic three-body term with parameter $\xi$. We use
this single parameter to reproduce the experiment at JILA \cite{ex4}.
For $\xi \ne 0$ in Eq. (\ref{d1}), ${\cal N}_{\mbox{norm}}
\ne 1.$

We solve the GP equation (\ref{d1}) numerically by time iteration with a
given initial solution. In the
time-evolution of the GP equation the radial and axial variables 
are dealt
with in independent steps. 
For this purpose
we
discretize 
it using time step $\Delta=0.001$ and space step $0.1$ for both 
$x$ and $y$ spanning $x$ from 0 to 15 and $y$ from $-25$ to 25. This 
domain of space was sufficient to encompass  the whole condensate wave
function even during and after explosion and collapse. The preparation of
the initial 
repulsive wave function is now a routine job and was done by increasing
the nonlinearity $n$ of the GP equation (\ref{d1}) by 0.0001 in each time
step $\Delta$ 
during time iteration starting with the known harmonic oscillator
solution of Eq. (\ref{d1}) for $n=\xi=0$ \cite{sk1}. 
The initial value of $n (=2.288)$   was attained after 22880 time steps.
The nonlinearity $n$ is then ramped from 2.288 to
$-9.805$ in 0.1 ms.
As one unit of dimensionless time $t$  is 0.009095 s, 0.1 ms corresponds
to 11  steps of time $\Delta$. In the present simulation, $n$ is  ramped
from 2.288 to $-9.805$ in the GP equation by equal amount in 11
steps.  The absorptive term $\xi$ was set equal
to zero
during above time iteration. 
Now the system is prepared for the simulation of collapse and
explosion.

For the remaining simulation the nonlinear term is maintained constant and
a nonzero value of $\xi$ is chosen. The time-evolution of the GP equation
is continued as a function of time $t=\tau_{\mbox{evolve}}$ starting at 0.  
Because of the nonzero dissipative term $\xi$, the normalization (\ref{5})
is no longer maintained, and the number of remaining atoms $N$ in the
condensate is given by $ N=N_0 {\cal N}_{\mbox{norm}}$, where $N_0$ is the
initial number. The time-evolution is continued using time step $\Delta
=0.001$.  After a small experimentation it is found that $\xi=3.12$ fits
the data of the experiment at JILA for $a_{\mbox{collapse}}=-30a_0$ $-$
their Fig. 1(b). This value of $\xi$ was used in all simulations reported
here.  The remaining number of atoms vs. time is plotted in Fig. 1.

It is pertinent to compare this value of $\xi (=3.12)$ with the
experimental measurement of three-body loss rate on $^{85}$Rb \cite{k3}.
For that, we need to restore the proper dimensions in the three-body term
of Eq. (\ref{d1}) by rewriting it as $i \xi (a N / l) ^2 |\psi|^4 l ^ 6 /
8$ and equating it to the convensional form $iK_3 N^2 |\psi|^4/ 2\omega$,
where $K_3=(4.24^{+0.70}_{-0.29}\pm 0.85)\times 10^{-25}$ cm$^6$/s is the
experimental rate \cite{k3}. From this we find $K_3=\xi a^2l^4\omega/4.$
Under experimental condition of an external magnetic field of 250 gauss on
$^{85}$Rb \cite{k3} the scattering length was $a \sim -350a_0$.
Consequently, the present value of $\xi (=3.12)$ corresponds to $K_3
\simeq 13 \times 10^{-25}$ cm$^6$/s, about three times larger than the
experimental value above.

In the experiment at JILA \cite{ex4} it was observed that the strongly
attractive  condensate after preparation remains stable with a constant
number of atoms 
for an interval of 
time $t_{\mbox{collapse}}$.
 This behavior is physically expected. Immediately after
the jump in scattering length from $7a_0$ to $-30a_0$, the attractive
condensate shrinks in size during $t_{\mbox{collapse}}$ until the central
density increases to a maximum. Then the absorptive three-body term takes
full control to initiate the sequence of 
collapse and explosion. Consequently, the number of atoms remains constant
for 
$\tau_{\mbox{evolve}}<t_{\mbox{collapse}}$. 
 The present result (full line) also shows a similar
behavior. However, in this simulation  the absorptive term is
operative from $\tau_{\mbox{evolve}}=0$ and the atom number decreases
right from beginning, albeit at a much smaller rate for
$\tau_{\mbox{evolve}}<t_{\mbox{collapse}}$. 

Donley et al. \cite{ex4} repeated their experiment with different 
values
of final 
scattering length.
Using 
the same value of $\xi (=3.12)$,
we also  repeated our calculation with the 
final scattering lengths: $a_{\mbox{collapse}}=-6.7a_0$ and $-250a_0.$
These results are also plotted in Fig. 1. The initial delay 
 $t_{\mbox{collapse}}$ in starting the explosion is large for small
$|a_{\mbox{collapse}}| $
 as we see in Fig. 1. After the initial delay, the decay rates of the
number of atoms are quite similar as observed in the experiment.  
Similar effect was observed in the experiment for an initial
condensate of 6000 atoms as shown in their Fig. 2.

After a sequence of collapse and explosion, Donley et al. \cite{ex4}
observed   a 
``remnant" condensate   of $N_{\mbox{remnant}}$ atoms at large times
containing a certain 
fraction 
of initial $N_0$ atoms. Experimentally, the fraction of atoms that went
into the remnant 
decreased with $|a_{\mbox{collapse}}|$ and was $\sim 40\%$ for
$|a_{\mbox{collapse}}|<10a_0$ and was $\sim 10\%$
$|a_{\mbox{collapse}}|>100a_0$.    Figure 1 also shows this behavior. 
We repeated the simulation of Fig. 1 with $N_0=6000$ where we also found a 
similar behavior. The number in the remnant
condensate could be much larger than $N_{\mbox{cr}}$.

The above evolution of the condensate after the jump in scattering length
to $-30a_0$ for $N_0=16000$ can be understood from a study of the wave
function and we display the central part of the wave function in Fig. 2
for
$\tau_{\mbox{evolve}}=0, 3.5, 4, $ and $10$ ms. The wave function
immediately after jump at time  $\tau _{\mbox{evolve}}=0$
is essentially the same as that before
the jump
at $-0.1$ ms. There is not enough time for the wave
function to get modified at $\tau_{\mbox{evolve}}=0$. From Fig. 2 we find
that at 3.5 ms the wave function is only slighty narrower than at 0 ms but
still smooth
and has
not yet collapsed significantly.
As $\tau _{\mbox{evolve}}$ increases, the wave
function contracts further and the explosion
starts. At 4 ms some spikes have appeared in the wave function showing the
beginning  of explosions and loss. From the study of the wavefunctions we
find that
the explosions start at $\tau _{\mbox{evolve}}=t_{\mbox{collapse}}\simeq
3.7 $ ms in agreement with the experiment at JILA. We
also find that at 3.7 ms
before the loss began the bulk BEC did not contract dramatically as also
observed in the experiment. In numerical simulation for this case we find
that 
at  $\tau _{\mbox{evolve}}=0, x_{\mbox{rms}}
=2.98 \mu$m and  $y_{\mbox{rms}}
=4.21 \mu$m  and 
at  $\tau _{\mbox{evolve}}=3.7$ ns, $x_{\mbox{rms}}
=2.53 \mu$m and  $y_{\mbox{rms}}
=4.10 \mu$m. From Fig. 2 we see that at
 10 ms the wave function is very spiky  
corresponding to many violent ongoing explosions.

%Donley et al. \cite{ex4} fitted the decay in the number of atoms to a
%decay constant
%$\tau_{\mbox{decay}}$  via the formula
%\begin{eqnarray} \label{dec}
%N(\tau_{\mbox{evolve}})&=&N_{\mbox{remnant}}+(N_0-N_{\mbox{remnant}})
%\nonumber \\
%&\times& 
%e^{(t_{\mbox{collapse}}-\tau_{\mbox{evolve}})/\tau_{\mbox{decay}}}
%\end{eqnarray}
%for $\tau_{\mbox{evolve}}>t_{\mbox{collapse}}$. They found that
%$\tau_{\mbox{decay}}$ is independent of $N_0$ and $a_{\mbox{collapse}}.$
%In Fig. 1 we also plot (dashed line) $N(\tau_{\mbox{evolve}})$ of
%Eq. (\ref{dec}) for
%$\tau_{\mbox{decay}}=2.8 {}$,
% $a_{\mbox{collapse}}=-30 a_0 $, and $N_{\mbox{remnant}}=5000.$   
% The results of the present simulation (full line) agree well
%with the average experimental result of Eq. (\ref{dec}) (dashed line).
%Our study with $N_0=6000$ also produced similar decay rates. 

Donley et al. \cite{ex4} also  observed that the remnant condensate
oscillated in a
highly
excited collective state with approximate  frequencies 
$2\nu _{\mbox{axial}}$ and  $2\nu _{\mbox{radial}}$   being predominantly
excited. The measured frequencies were 
13.6(6)
Hz
and 33.4(3) Hz. To
find if
this behavior emerges from the present simulation we plot  in Fig. 3
 sizes $x_{\mbox{rms}}$ and  $y_{\mbox{rms}}$ vs. time for the 
condensate after the jump in the scattering length to $-30a_0$ from $7a_0$
for $N_0=16000 $. 
 Excluding the first 20
ms when the  remnant  condensate is being formed, we find a periodic
oscillation 
in    $x_{\mbox{rms}}$ and  $y_{\mbox{rms}}$ with frequencies
      34      Hz   and  13.6    Hz,  respectively.    In addition, for
$x_{\mbox{rms}}$,  
superposed on 34 Hz  we note a prominent frequency of about 3.5 Hz, e.g., 
$0.2\nu _{\mbox{radial}}$.

Though we have explained some aspects of the experiment at JILA,
there are certain detailed features that have not been obtained in this
Letter. Donley et al. \cite{ex4} have classified the emitted atoms in
three
categories: burst,
jet and missing atoms. Detailed study of the wave function is
needed to identify the jet atoms that appear possibly from the spikes in
the wave function when the collapse is suddenly interrupted. The jets were
also not always found to be symmetric about the condensate axis
\cite{ex4}. This makes an axially symmetric mean-field model as used in
this Letter inappropriate
for a description of the jet atoms. Moreover a clear-cut distinction
between the burst and missing atoms emitted during explosions seems to be
difficult in the present model.  Also because of the missing (undetected)
atoms it is difficult to predict the energy distribution of the burst
atoms in a mean-field analysis which will only yield the total energy
distribution of the burst plus missing atoms.  A careful analysis of the
energy of the emitted atoms is required for explaining these detailed
features and would be a welcome future theoretical investigation.

In conclusion, we have employed a numerical simulation based on the
accurate
solution \cite{sk1}
of the mean-field Gross-Pitaevskii equation with a cylindrical
trap to study the dynamics of collapse and explosion as observed in the
recent experiment at JILA \cite{ex4}. In the GP equation we include a
quintic
three-body nonlinear recombination loss term which accounts for the decay
of the strongly attractive condensate.  The results of the present
simulation accounts for the essential features of that xperiment.
We find that the exact inclusion of the asymmetry parameter $\lambda$ 
in the simulation is essential for a proper description of the 
experiment. 
In the  experiment at JILA a strongly attractive $^{85}$Rb condensate was
prepared by ramping the scattering length to a large negative value and
the subsequent decay of the condensate was measured. We have been able to
understand the following features of this dynamics from the present
numerical simulation: (1) The condensate undergoes a sequence of collapse
and explosion and finally stabilizes to a remnant condensate at
large times containing about $\sim 10\% - 40\%$ of the initial number 
$N_0$. The number in the remnant condensate can
be larger than the critical number for collapse $N_{\mbox{cr}}$ for the 
same atomic 
interaction.  (2) Both in the experiment and our simulation the 
remnant condensate executes radial and axial oscillations in a highly
excited collective state for a long time  with frequencies 
$2\nu_{\mbox{radial}}$ and $2\nu_{\mbox{axial}}.$ 
(3) The rate of decay of the number of particles is
reasonably constant independent of $N_0$ and $a_{\mbox{collapse}}.$  
(4)
After the sudden change in the scattering length to a large negative
value,
the condensate needs an interval of time $t_ {\mbox{collapse}} $ before it
experiences loss via explosions.
Consequently, the decay starts after the interval of time $t_
{\mbox{collapse}}$.

The investigation  is supported in part by the CNPq and FAPESP 
of Brazil.

\vskip 1cm

{\bf Figure Caption:}

1. Number of remaining atoms in the $^{85}$Rb condensate of 16000 atoms
after ramping the scattering length from $a_{\mbox{initial}}=
7a_0$ to $a_{\mbox{collapse}}=
 -6.7a_0, -30a_0$ and $-250a_0$ in 0.1
ms as
a
function of evolution time $\tau_{\mbox{evolve}}$ in ms. 
The solid circles \cite{ex4}
are from experiment with $a_{\mbox{collapse}} =-30a_0$   and the
theoretical curves are labeled by the values of $a_{\mbox{collapse}}$.
%Average over
%experimental results from Eq. (\ref{dec}) is also plotted (dashed line).

2. The central part of the dimensionless wave function $|\phi(x,y)|\equiv
|\varphi(x,y)/x|$
of the condensate on $0.1\times 0.1$ grid 
after the jump
in the scattering
length of a BEC of  16000 $^{85}$Rb atoms from $7a_0$ to
$a_{\mbox{collapse}}=-30a_0$ at times $\tau_{\mbox{evolve}}
=0$, 3.5 ms,  4 ms, and 10 ms. 

3. The dimensionless rms sizes $x_{\mbox{rms}}$ (full line) and
$y_{\mbox{rms}}$
(dashed line) expressed in units of $l/\sqrt 2$ 
after the jump
in the scattering
length of a BEC of  16000 $^{85}$Rb atoms from $a_{\mbox{initial}}=7a_0$
to
$a_{\mbox{collapse}}=-30a_0$
as functions of
time  $\tau_{\mbox{evolve}}$.

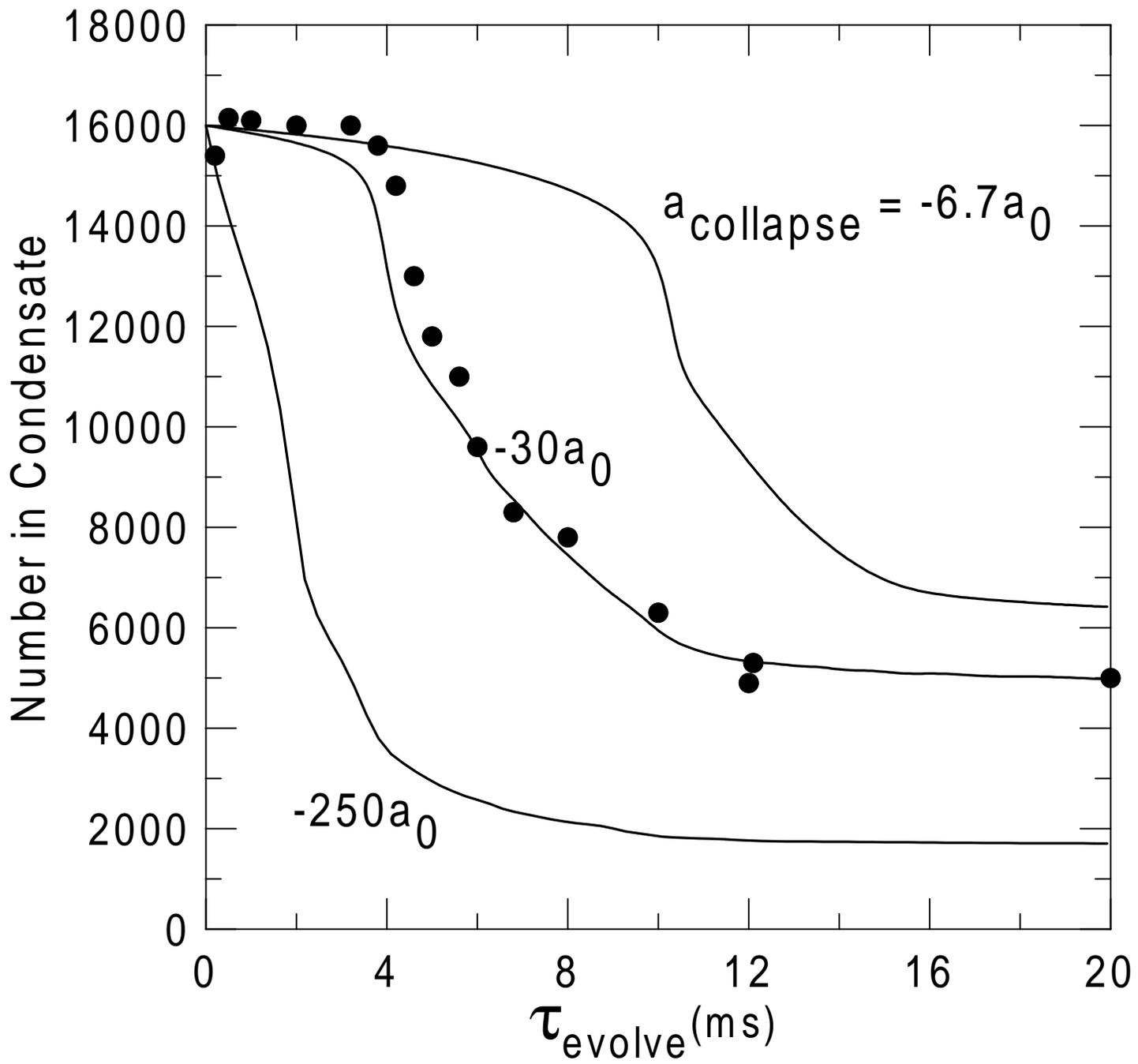
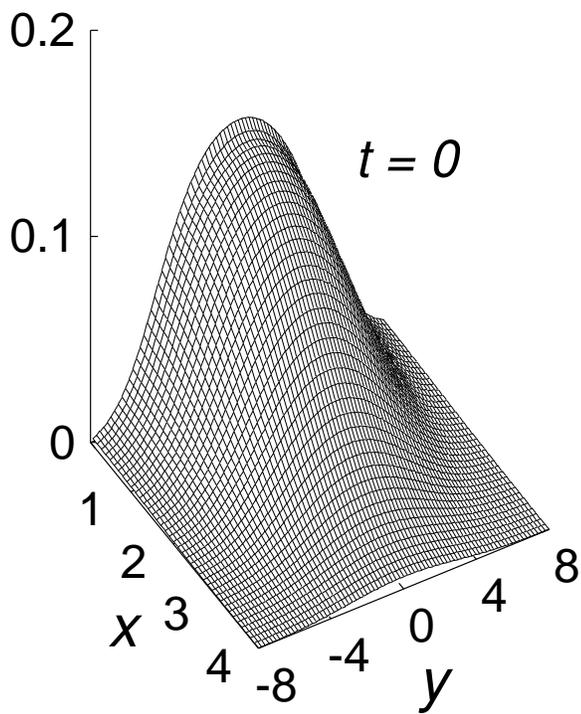
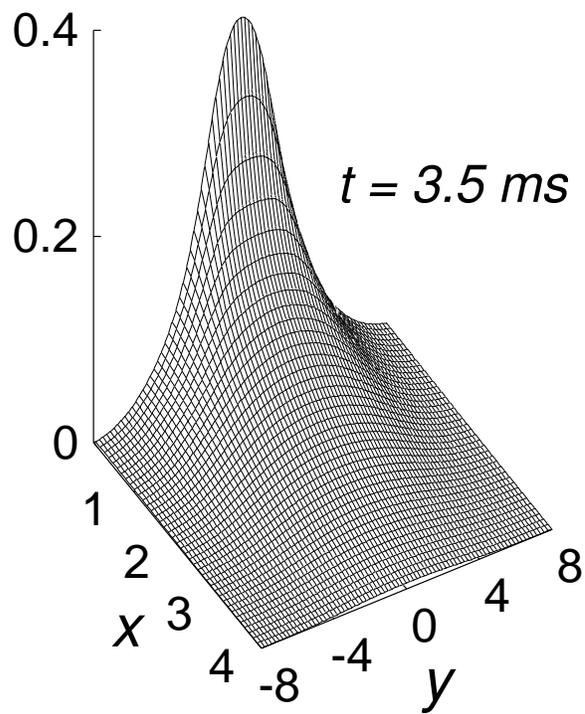
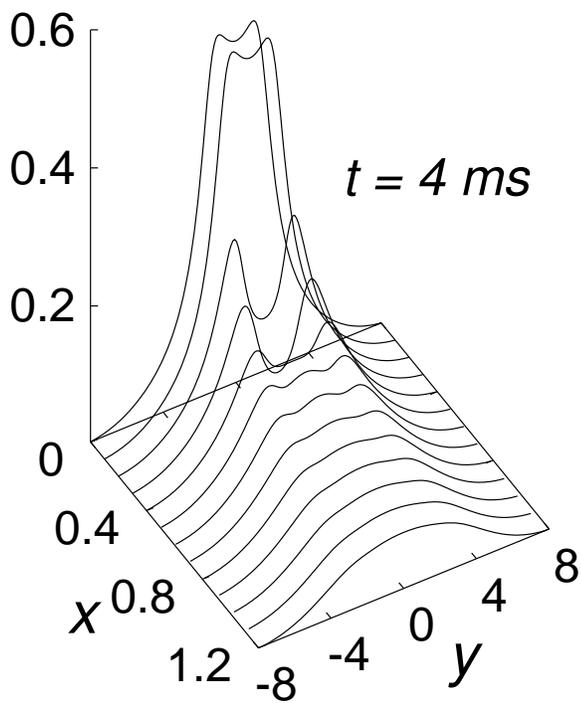
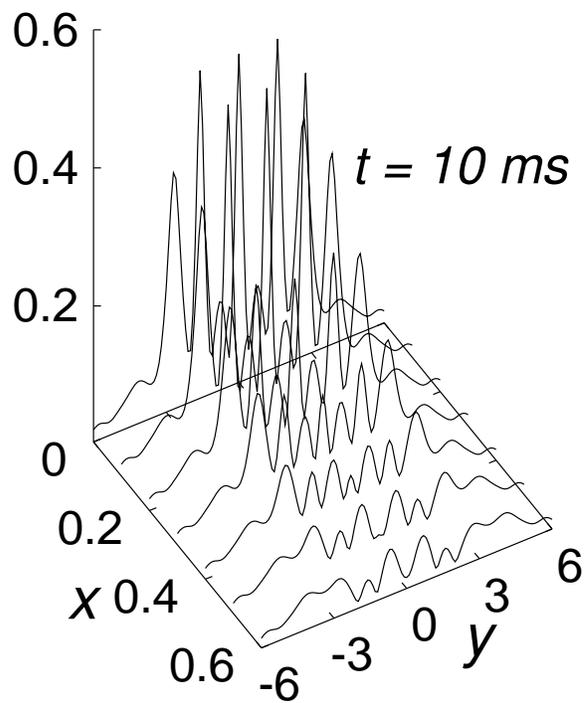
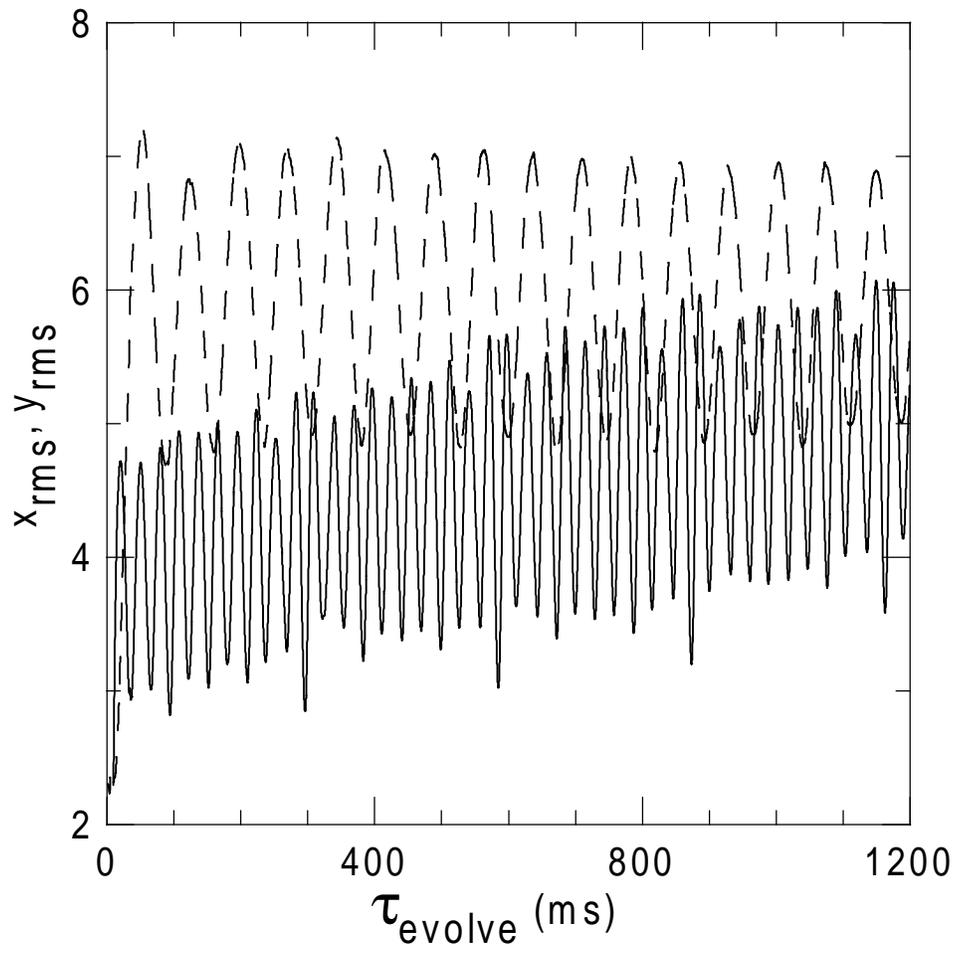
\end{document}